# Pressure-Induced Superconductivity in Layered Transition-metal Chalcogenides (Zr,Hf)GeTe$_4$ Explored by Data-driven Approach

Ryo Matsumoto[a], Zhufeng Hou[c], Shintaro Adachi[b,d], Sayaka Yamamoto[b,e], Hiromi Tanaka[f], Hiroyuki Takeya[b], Tetsuo Irifune[g], Kiyoyuki Terakura[h] and Yoshihiko Takano[b,e]

[a]*International Center for Young Scientists (ICYS),*
*National Institute for Materials Science, Tsukuba, Ibaraki 305-0047, Japan*
[b]*International Center for Materials Nanoarchitectonics (MANA),*
*National Institute for Materials Science, Tsukuba, Ibaraki 305-0047, Japan*
[c]*State Key Laboratory of Structural Chemistry, Fujian Institute of Research on the Structure of Matter, Chinese Academy of Sciences, Fuzhou 350002, China*
[d]*Kyoto University of Advanced Science, Kyoto 615-8577, Japan*
[e]*University of Tsukuba, Tsukuba, Ibaraki 305-8577, Japan*
[f]*National Institute of Technology, Yonago College, Yonago, Tottori 683-8502, Japan*
[g]*Geodynamics Research Center, Ehime University, Matsuyama, Ehime 790-8577, Japan*
[h]*National Institute of Advanced Industrial Science and Technology, Tsukuba, Ibaraki 305-8560, Japan*

**Abstract**

Layered transition-metal chalcogenides (Zr,Hf)GeTe$_4$ were screened out from database of Atomwork as a candidate for pressure-induced superconductivity due to their narrow band gap and high density of state near the Fermi level. The (Zr,Hf)GeTe$_4$ samples were synthesized in single crystal and then the compositional ratio, crystal structures, and valence states were investigated via energy dispersive spectrometry, single crystal X-ray diffraction, and X-ray photoelectron spectroscopy, respectively. The pressure-induced superconductivity in both crystals were first time reported by using a diamond anvil cell with a boron-doped diamond electrode and an undoped diamond insulating layer. The maximum superconducting transition temperatures of ZrGeTe$_4$ and HfGeTe$_4$ were 6.5 K under 57 GPa and 6.6 K under 60 GPa, respectively.

# 1. Introduction

After the discovery of superconductivity in elemental Hg with transition temperature ($T_c$) of 4.2 K by Heike Kamerling Onnes in 1911, explorations for room temperature superconductors have been continued by a lot of researchers around the world. There were some important breakthroughs in the history of explorations for superconductors, for example, the cuprate superconductors [1], Fe-based superconductors [2], and hydrogen-rich superconductors [3]. Among them the cuprate superconductor of $(La,Ba)_2CuO_4$ and the Fe-based superconductor of $La(O,F)FeAs$ were discovered by serendipity through the study of the dielectric materials and the transparent semiconducting materials, respectively. After the experimental discoveries of these materials, a lot of theoretical studies were developed to explain their mechanisms of the high-$T_c$ superconductivity.

On the other hand, some theoretical studies using the first-principles calculations had predicted the high-$T_c$ superconductivity in hydrogen-rich compounds in advance [4], thereafter, the record $T_c$ at 203 K was reported experimentally in $H_3S$ [3,5,6]. The theoretically calculated $T_c$ in $H_3S$ well coincided with the experimentally reported value [7]. Moreover, two group of high-pressure physics reported extremely high-$T_c$ above 250 K in $LaH_x$ [8,9]. The superconductivity in $LaH_x$ was also predicted by the theoretical studies before these experiments [10].

Furthermore, the idea of using informatics techniques in conjunction with the data-driven approach based on high-throughput computation for functional materials design, namely materials informatics, recently has been carried out in practice and also has produced remarkable outcomes [11-16]. However, the exploration for the inorganic material, especially superconductors via the data-driven approach is still minority. Our group recently performed the data-driven approach to explore new pressure-induced superconductors using a database and the density functional theory (DFT) calculations [17-19]. The candidate compounds were screened out by focusing a "flat band" near a Fermi level such as multivalley [20], pudding mold [21], and topological-type [22] structures. If the flat band approaches to the Fermi level, a high density of state (DOS) would be advantageous for superconductivity. For example, an existence of singularity of DOS, known as a van Hove singularity (vHs) was predicted near the Fermi level in compressed high-$T_c$ $H_3S$ [7], A15 type compounds [23], and high-$T_c$ cuprates [24]. Indeed, new pressure-induced superconductivity was discovered in real compounds of $SnBi_2Se_4$ [17], $PbBi_2Te_4$ [18], and $AgIn_5Se_8$ [19] through the high-throughput screening.

Among the candidate compounds from the screening, $(Zr,Hf)GeTe_4$ are significantly focused from a viewpoint of their unique crystal structure. The compounds were first synthesized in single crystal of hair-like fibers in previous report [25]. If (Zr,Hf) and Ge are considered as a cation atom, $(Zr,Hf)GeTe_4$ can be written by $MTe_2$ ($M$ = Hf, Zr), as transition metal dichalcogenides (TMDs). One of the practical issues of TMDs is their atomically flat passivated surface that induces a poor adhesion between an electrodes and other surrounding materials due to the weak interaction. On the other hand, since $(Zr,Hf)GeTe_4$ has a zigzag surface which brings a larger surface area than that of flat one, an improvement of the week interaction could be expected [26].

In this paper, we first report a pressure-induced superconductivity in $(Zr,Hf)GeTe_4$ explored by the data-driven approach. The $ZrGeTe_4$ and $HfGeTe_4$ were synthesized in single crystals with hair-like fiber shape. The crystal structure, compositional ratio, and valence state of the sample crystals were analyzed by the X-ray diffraction (XRD), energy dispersive X-ray spectrometry (EDX), and X-ray photoelectron spectroscopy (XPS), respectively. The electrical resistance of the samples was evaluated under high pressure using a diamond anvil cell (DAC) with a boron-doped diamond electrode and an undoped diamond insulating layer [27-30].

## 2. Electronic band structures of (Zr,Hf)GeTe$_4$

The electronic band structures of the selected materials (Zr,Hf)GeTe$_4$ were calculated under ambient pressure and 10 GPa. The details of our screening scheme in the high-throughput first-principles calculations were given in our previous paper [17]. Figure 1 (a) shows the band structures and the total density of states (DOS) of ZrGeTe$_4$ at ambient pressure and (b) under high pressure of 10 GPa, which were obtained by first-principles calculations within the generalized gradient approximation. The band structure of ZrGeTe$_4$ at ambient pressure exhibits a narrow gap of 0.40 eV and large DOS near the Fermi level. The band gap is rapidly closed by applying pressure up to 10 GPa, and then ZrGeTe$_4$ exhibits a metallic feature. The band structure and DOS of HfGeTe$_4$ at ambient pressure and under high pressure of 10 GPa are displayed in Fig. 1 (c,d). Although GeHfTe$_4$ shows similar electronic states with those of ZrGeTe$_4$, its band gap is slightly wider. It is also noted that the conduction band edges of (Zr,Hf)GeTe$_4$ are contributed primarily by the (4$d$, 5$d$) orbitals of (Zr,Hf) and the 5$p$ orbital of Te-Te dimer in (Zr,Hf)GeTe$_4$. The slightly wider band gap of HfGeTe$_4$ might be because Hf 5$d$ orbital is slightly shallower than the Zr 4$d$ orbital. Since the experimental study reported the wider band gap in HfGeTe$_4$ as compared with ZrGeTe$_4$, our calculation results are reasonable.

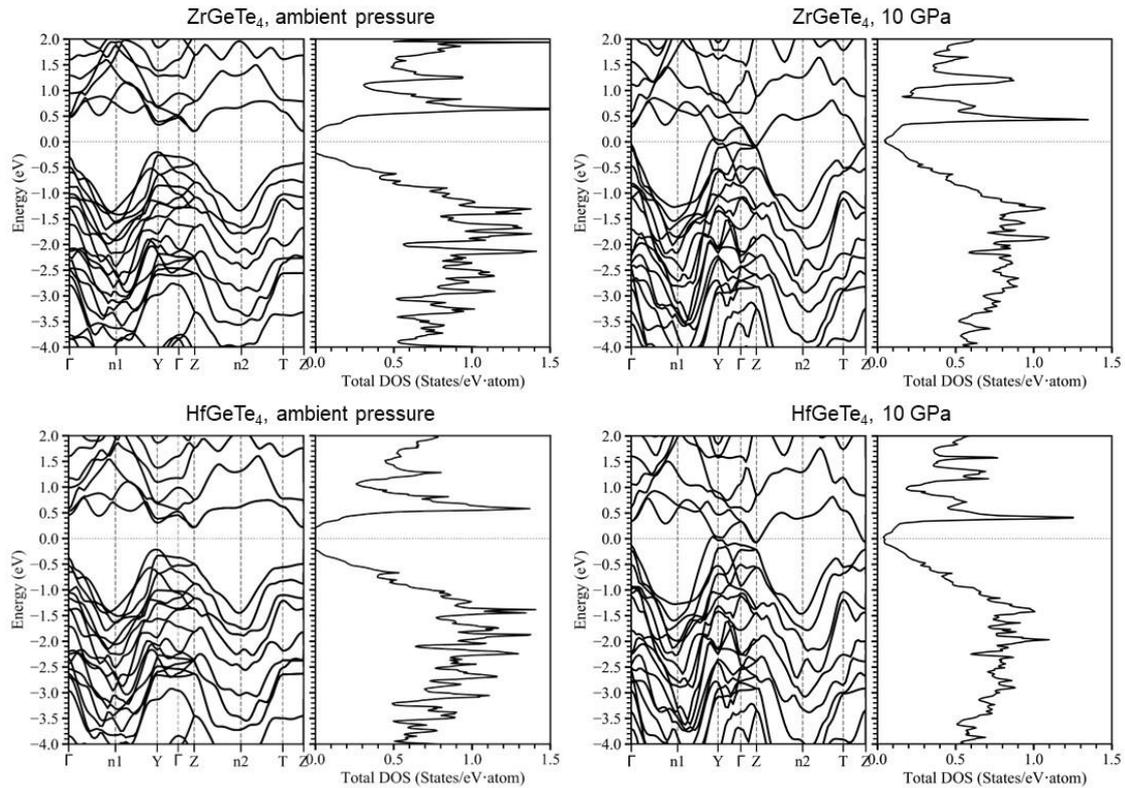

**Figure 1. Electronic band structures and total density of states (DOS) of (Zr,Hf)GeTe$_4$ under ambient pressure and 10 GPa, obtained by the first-principles calculations within the generalized gradient approximation.**

# 3. Experimental procedures

## 3.1 Sample synthesis

Single crystals of (Zr,Hf)GeTe$_4$ were prepared by referring the previous report [25]. Starting materials of Zr or Hf grains (99.98%), Ge powders (99.99%), and Te chips (99.999%) were put into an evacuated quartz tube in stoichiometric compositions of (Zr,Hf)GeTe$_4$. The ampoules were heated at 650ºC for 20 hours, then subsequently at 900ºC for 50 hours, slowly cooled to 500ºC for 50 hours followed by furnace cooling.

## 3.2 Characterization

The crystal structures of the grown samples were determined by a single crystal XRD by use of the XtaLAB mini (Rigaku) with Mo-K$\alpha$ radiation ($\lambda$ = 0.71072 Å) and the program SHELXT on the WinGX software and refined using the program ShelXle [31-33]. The software VESTA [34] was used for a depiction of the solved crystal structures. The chemical composition was analyzed by the EDX, using the JSM-6010LA (JEOL). The valence states were estimated by a peak separation for the core level XPS spectra using the AXIS-ULTRA DLD (Shimadzu/Kratos) with monochromatic Al K$\alpha$ X-ray radiation ($h\nu$ = 1486.6 eV). The surface of samples was milled in the main chamber with high vacuuming pressure of the order of $10^{-9}$ Torr using an Ar gas cluster ion beam (GCIB) with 20 keV beam energy before acquiring the XPS spectra. The mean size of one cluster was approximately 1000 atoms, the scanning area of the GCIB was about 2 mm$^2$, and the beam current was about 20 nA. The GCIB radiation provides a slow etching for the sample surface without a change of the intrinsic chemical state [35]. The acquired spectra were analyzed through a background subtraction by active Shirley algorism and a pseudo-Voigt function peak fitting on the COMPRO software [36].

## 3.3 Configuration of high-pressure cell

In the electrical resistance measurements of (Zr,Hf)GeTe$_4$ under high pressure, we used originally designed DAC equipping the boron-doped diamond (BDD) electrodes and the undoped diamond (UDD) insulating layer [37]. Figure 2 (a) shows a schematic image of the DAC configuration for the measurements of ZrGeTe$_4$. The cleaved sample was placed at the BDD electrodes on the center of a nano-polycrystalline diamond anvil [38]. The electrodes and a metal gasket were separated by UDD insulation. In the measurements for HfGeTe$_4$, culet-type anvil with BDD electrodes and UDD insulating layer was used. The details of the cell configuration were described elsewhere [37]. Figure 2 (b) shows a microscope image for ZrGeTe$_4$ on the high-pressure cell. The mixture powder of cubic boron nitride and ruby manometer were used as a pressure-transmitting medium. The applied pressure was detected by a pressure-driven peak shift of ruby fluorescence [39] and a Raman mode of the diamond anvil [40] using the inVia Raman Microscope (RENISHAW). The electrical resistance was measured by a four-terminal method using a physical property measurement system (Quantum Design: PPMS).

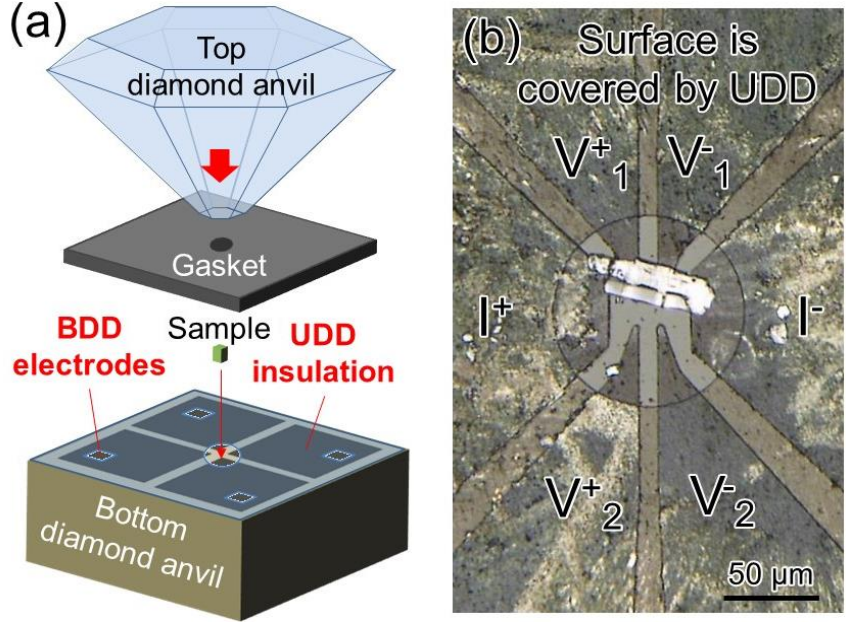

**Figure 2.** (a) Schematic image for originally designed diamond anvil cell (DAC) with boron-doped diamond (BDD) electrodes and undoped diamond (UDD) insulating layer. (b) Sample space of bottom diamond anvil with ZrGeTe$_4$ single crystal.

## 4. Results and discussion

4.1 Composition and structural analysis

Figure 3 shows the SEM images of the obtained (a) ZrGeTe$_4$ and (b) HfGeTe$_4$ crystals. As reported in the previous study [25], the crystals formed as hair-like fibers. The chemical composition of the obtained ZrGeTe$_4$ single crystal is Zr : Ge : Te = 0.99 : 0.99 : 4 from the EDX analysis, which is consistent with the nominal composition of ZrGeTe$_4$. On the other hand, the estimated compositional ratio of HfGeTe$_4$ is Hf : Ge : Te = 0.89 : 0.99 : 4, which suggests a Hf deficiency.

The single crystal structural analysis of (Zr,Hf)GeTe$_4$ were successfully performed. The refinement for ZrGeTe$_4$ was converged to the $R_1$ values of 4.04% for all data. The compound was supposed to be indexed by orthorhombic unit cell (space group: $Cmc2_1$) with lattice constants of $a$ = 3.986(6) Å, $b$ = 15.95(3) Å and $c$ = 11.021(17) Å. The variation of the occupancy of a specific site made no significant reduction of $R_1$ values, suggesting no deficiencies at each site in obtained ZrGeTe$_4$. The crystal structure of HfGeTe$_4$ was also assigned as an orthorhombic lattice (space group: $Cmc2_1$) with lattice constants of $a$ = 3.9892(17) Å, $b$ = 15.972(7) Å and $c$ = 10.982(4) Å. On the other hand, the $R_1$ value in the refinement of HfGeTe$_4$ was significantly reduced from 6.53% to 5.76% as the occupancy of a Hf site was varied, indicating that the obtained HfGeTe$_4$ have deficiency at the Hf site, corresponding to the previous report for the synthesis [41]. The compositional ratio was Hf$_{0.83}$GeTe$_4$ from the refinement, which is well agreement with the EDX results. Although the band gap of HfGeTe$_4$ is slightly higher than that of ZrGeTe$_4$ from calculations, the carrier concentration of HfGeTe$_4$ would be higher than that of ZrGeTe$_4$ due to the Hf deficiency. It can be interested that the differences between ZrGeTe$_4$ and HfGeTe$_4$ onto the electrical transport properties reflecting their deficiency features.

Figure 3 (c) displayed the schematic images of the crystal structures of ZrGeTe$_4$ and (d) HfGeTe$_4$ based on the single crystal structural analysis. The inserted optical images are picked up samples on the glass capillary for the single crystal XRD measurement. The isostructural compounds (Zr,Hf)GeTe$_4$ demonstrate the nature of the van der Waals gap separated layers [25]. The anisotropic structure is constructed by one dimensional chain-like structures with trigonal prisms. According to a precise Raman spectroscopy investigation, ZrGeTe$_4$ exhibits highly in-plane anisotropic two-dimensional feature, suggesting its potential application in nano-electronic devices [42].

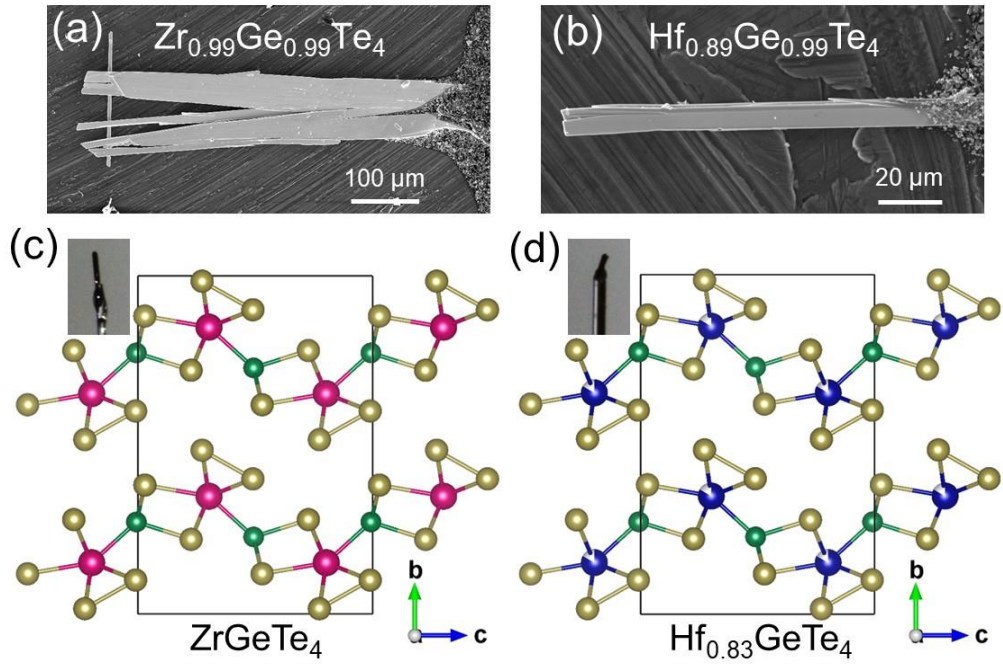

**Figure 3. (a) SEM images of ZrGeTe$_4$ and (b) HfGeTe$_4$. (c) Schematic images of the crystal structures of ZrGeTe$_4$ and (d) HfGeTe$_4$ based on the single crystal structural analysis. The inserted optical images are picked up samples on the glass capillary for the single crystal XRD measurement.**

4.2 Valence state

The valence states of Zr in ZrGeTe$_4$ and Hf in HfGeTe$_4$ were investigated by XPS. Figure 4 (a) shows Zr 3$d$ core-level spectrum of ZrGeTe$_4$. The spectrum was deconvoluted into four peaks as labeled in the fig. 4 (a). The main components of peak 1 at 185.0 eV and peak 2 at 182.6 eV are corresponding to a Zr$^{4+}$ valence state with a spin–orbit splitting of 2.4 eV [43]. The minor components of peak 3 at 182.9 eV and peak 4 at 180.5 eV are originated from a Zr$^{2+}$ state because a energy shift of 2.1 eV between Zr$^{4+}$ and Zr$^{2+}$ was accordingly reported by previous study [43]. Figure 4 (b) shows Hf 4$f$ core-level spectrum of HfGeTe$_4$. The Hf 4$f$ orbital also exhibited the two kinds of valence states. The peak 1 at 19.2 eV and the peak 2 at 17.6 eV are corresponding to a Hf$^{4+}$ valence state with a spin–orbit splitting of 1.6 eV, and the peak 3 at 17.5 eV and the peak 4 at 16.0 eV are from Hf$^{2+}$ state [44]. These features of the valence fluctuation of Zr and Hf are consistent with predicted electronic structures from molecular orbital calculations in previous study [25].

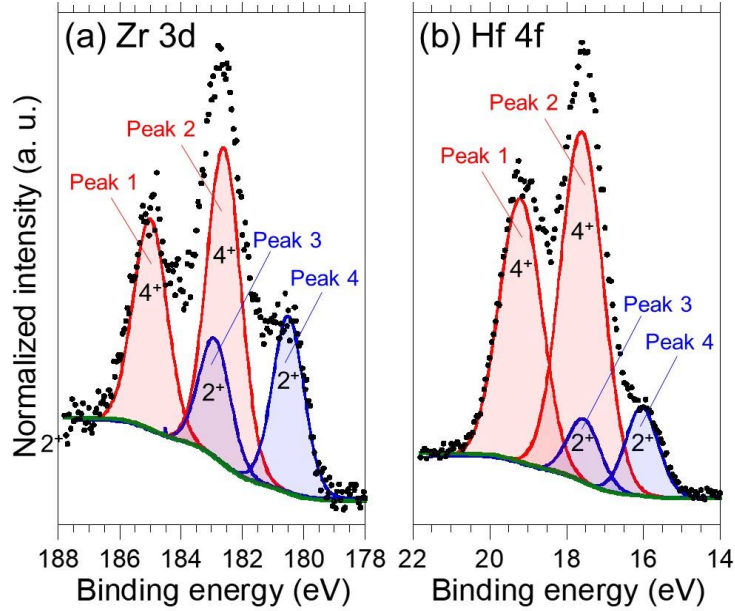

**Figure 4. (a) Zr *3d* core-level spectrum of ZrGeTe$_4$ and (b) Hf *4f* core-level spectrum of HfGeTe$_4$.**

4.3 Electrical resistance measurement of (Zr,Hf)GeTe$_4$ under high pressure

Figure 5 (a) shows temperature dependence of resistance in ZrGeTe$_4$ under various pressures from 8.1 GPa to 30.3 GPa. The d$R$/d$T$ of ZrGeTe$_4$ under 8.1 GPa was negative from 300 K to 200 K, indicating a semiconducting characteristic. The slope of the resistance curve tended to metallic below 200 K, and returned to semiconducting below 50 K. This complex behavior of the resistance curve is consistent with that of the ambient property reported by previous study [25]. The absolute value of resistance at 300 K was drastically decreased with an increase of the pressure, and then the decrease was saturated at 17.4 GPa. The semiconducting behavior at around 300-200 K and below 50 K was changed to metallic feature with increase of the applying pressure. Instead of the suppression of the semiconducting features and the saturation of the resistance decreasing, a sudden drop of the resistance appeared at low temperature at 17.4 GPa. Figure 5 (b) shows the enlargement plots of the resistance curve below 10 K. The resistance started to decrease from around 2 K, indicating an onset temperature ($T_c^{onset}$) of a pressure-induced superconductivity in ZrGeTe$_4$. The superconducting transition was observed more clearly under 23.7 GPa with a zero resistance at around 2 K. Moreover, a multi-step transition was observed from 23. 7 GPa with drastic enhance of $T_c^{onset}$, suggesting an emergence of a second superconducting phase. The transition temperature at zero resistance ($T_c^{zero}$) was increased up to 2.4 K under 30.3 GPa.

Such an anomaly under the pressure below 17.4 GPa at normal state of the temperature dependence of resistance was found in various 2D layered materials, ZrTe$_3$ [45], HfTe$_3$ [46], 1$T$-TiSe$_2$ [47], and so on [48], as a sign of charge density wave (CDW). Especially the high-pressure study on 1$T$-TiSe$_2$ revealed that the anomaly suppressed by increase of the applied pressure, and then the superconductivity was driven [47]. A coexistence of superconductivity and CDW was also reported in HfTe$_3$ [46]. The detailed analysis of the possible existence of CDW in this compound and of its relationship with the pressure-induced superconductivity will be leaved in our future work.

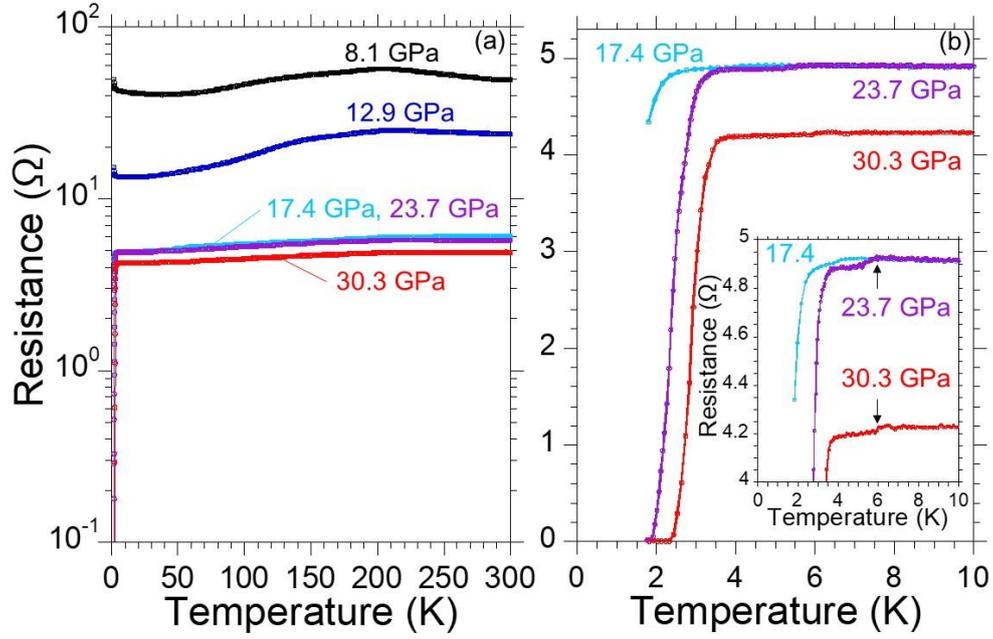

**Figure 5.** (a) Temperature dependence of resistance in ZrGeTe$_4$ under various pressures from 8.1 GPa to 30.3 GPa. (b) Enlargement plots below 10 K. The inset shows more enlarged plots to see the onset of superconducting transition.

Figure 6 (a) shows temperature dependence of resistance in ZrGeTe$_4$ under the pressures from 23.7 GPa to 80 GPa. The absolute value of the resistance at 300 K continued to decrease by increasing the pressure up to 80 GPa. The behavior of the resistance curve much tended to metallic under higher-pressure region. A resistance temperature exponent $n$ was obtained through a fitting procedure from 7 K to 50 K under 57 GPa, that is derived from the standard resistance fit, $R(T) = R_0 + AT^n$, where $R_0$ is the residual resistance, $A$ is a characteristic constant and $T$ is the temperature. The inset of Fig. 6 (a) shows the temperature dependence of resistance under 57 GPa below 50 K with the fitting curve. The resistance exponent $n$, residual resistance $R_0$ and characteristic constant $A$ are found to be 1.96, 3.43 $\Omega$ and 2.22×10$^{-5}$ $\Omega$/K$^2$, respectively. The resistance exponent $n \sim 2$ is suggestive of a pure electronic correlation-dominated scattering mechanism [49]. Figure 6 (b) shows the enlargement plots for temperature dependence of resistance in ZrGeTe$_4$ under the pressure below 8 K. Although the signal of $T_c^{\mathrm{onset}}$ was tiny, it was maximumly enhanced up to ~6.5 K at 57 GPa. Further application of pressure gave slight decrease of the $T_c^{\mathrm{onset}}$. The appearance of higher $T_c$ phase in small drop of resistance in ZrGeTe$_4$ was further corroborated by the resistance curve in applied magnetic fields. As seen from Fig. 6 (c), the temperature of the small drop of resistance gradually shifted toward lower temperatures with increasing magnetic fields. This fact suggests that the small drop of resistance is originated from the superconductivity, namely a coexistence of two kinds of superconducting phase in the compressed ZrGeTe$_4$. After an increase of the pressure above 80 GPa, the bottom diamond anvil itself including the electrodes was cracked, and then we decreased the pressure.

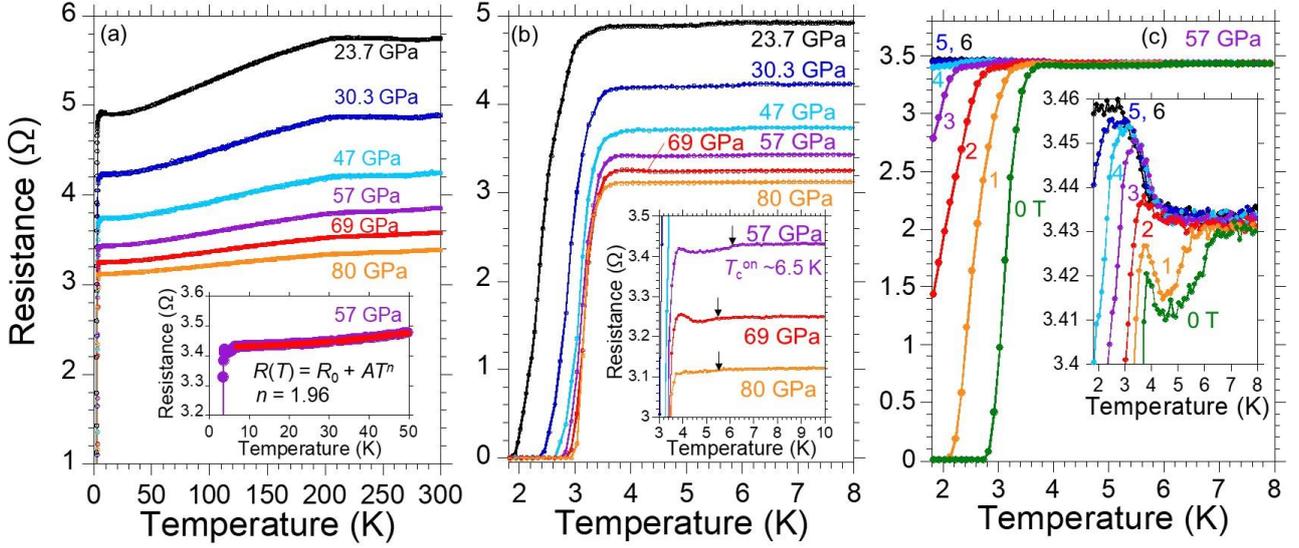

**Figure 6. (a)** Temperature dependence of resistance in ZrGeTe$_4$ under the pressures from 23.7 GPa to 80 GPa. The inset shows enlargement below 50 K at 57 GPa with a fitting curve of $R(T) = R_0 + AT^n$. **(b)** Enlargement plots for temperature dependence of resistance in ZrGeTe$_4$ under pressure below 8 K. The inset shows more enlarged plots to see the onset of superconducting transition. **(c)** Temperature dependence of resistance in ZrGeTe$_4$ at 57 GPa under various magnetic fields up to 6 T. The inset shows the enlargement around the small drop of resistance.

Figure 7 (a) shows temperature dependence of resistance in HfGeTe$_4$ under various pressures from 3.2 GPa to 15.5 GPa. As same as the ZrGeTe$_4$ case, HfGeTe$_4$ exhibited semiconducting behavior with an anomaly at around 200 K. The absolute value of the resistance at 300 K was continued to decrease against an increase of the pressure up to 13.9 GPa with a suppression of the anomaly. A sudden drop of the resistance from 3 K corresponding to the $T_c^{onset}$ of a pressure-induced superconductivity in HfGeTe$_4$ was observed at 8.1 GPa. A slight increase of the resistance just before the superconducting transition, as shown in the enlarged plots of Fig. 7 (b), could be suppressed by applying magnetic fields, indicating a sign of precursory phenomenon of superconductivity. The $T_c^{onset}$ was enhanced by an increasing pressure, and clear zero-resistance was observed at $T_c^{zero} = 2$ K under 13.9 GPa. Although the decrease of the resistance at 300 K was almost saturated at 15.5 GPa, the $T_c$ enhancements were continued.

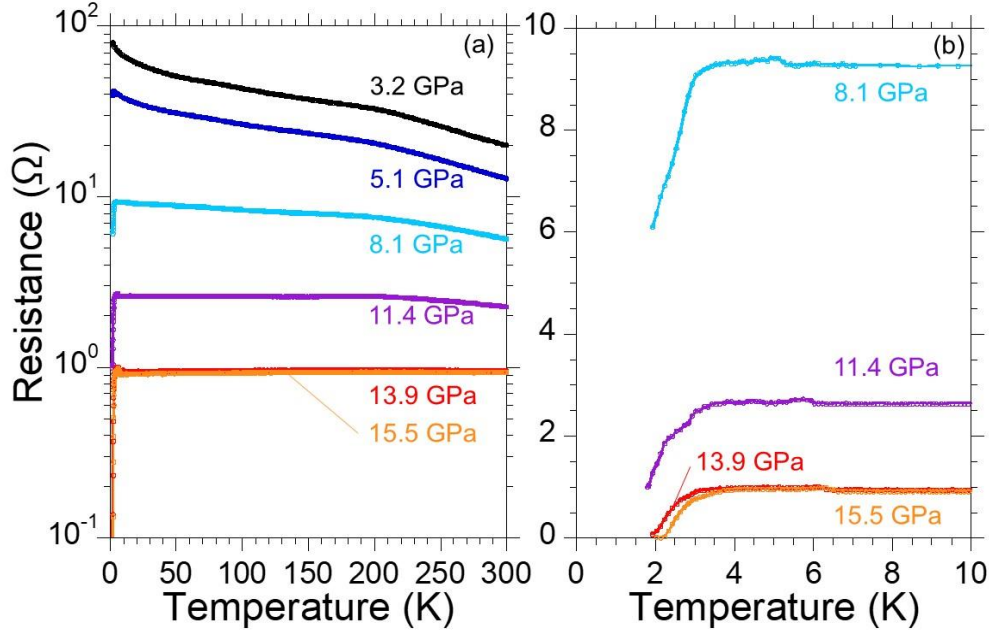

**Figure 7.** (a) Temperature dependence of resistance in HfGeTe$_4$ under various pressures from 3.2 GPa to 15.5 GPa. (b) Enlarged plots below 10 K.

Figure 8 (a) shows temperature dependence of resistance in HfGeTe$_4$ under various pressure from 15.5 GPa to 101 GPa. A slope of the resistance d$R$/d$T$ completely became positive at 86 GPa. The inset of Fig. 8 (a) shows the temperature dependence of resistance under 86 GPa below 50 K with the fitting curve of $R(T) = R_0 + AT^n$ with the resistance exponent $n$ of 2.75, residual resistance $R_0$ of 1.3 and characteristic constant $A$ of $4.54 \times 10^{-8}$ $\Omega$/K$^2$. The $n>2$ feature indicates a deviation from a pure electronic correlation-dominated scattering mechanism, namely a contribution from an interband electron-phonon scattering, as observed in Nb(S,Se)$_2$ ($n$=3) [50] ZrSiS ($n$ = 3) [49], LaSb ($n$ = 4) [51] and so on [52]. Figure 8 (b) shows the enlarged plots for temperature dependence of resistance below 8 K in HfGeTe$_4$ under the pressures. The slight increase of the resistance just before the superconducting transition could be still observed under 32 GPa. On the other hand, the higher $T_c^{onset}$ suddenly appeared at around 6.6 K instead of a suppression of the anomaly. The signal of the resistance drop from higher $T_c^{onset}$ was more emphasized by increase of the pressure to 60.2 GPa. After the $T_c^{onset}$ became maximum at 6.8 K under 60.2 GPa, it decreased under further compression up to 101 GPa. Figure 8 (c) shows a temperature dependence of resistance in HfGeTe$_4$ at 60.2 GPa under various magnetic fields up to 6 T. The inset shows the enlargement around the superconducting transition. Both the higher and lower $T_c$ value were shifted to lower temperature side against the applied magnetic field. This feature suggests that HfGeTe$_4$ exhibits the two kinds of the pressure-induced superconductivity with different $T_c$ value as same as ZrGeTe$_4$ case. It is worthy to note that the BDD electrodes used for the measurements were not broken after the pressure-reduction from 101 GPa to ambient pressure.

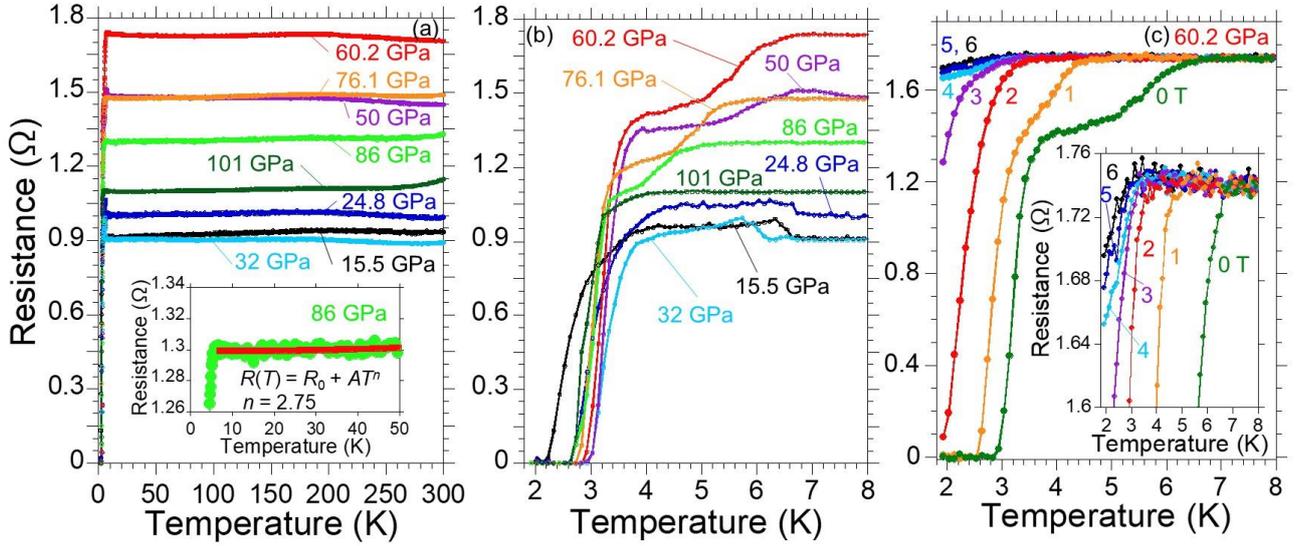

**Figure 8.** (a) Temperature dependence of resistance in HfGeTe$_4$ under the pressures from 15.5 GPa to 101 GPa. The inset shows enlargement below 50 K at 86 GPa with a fitting curve of $R(T) = R_0 + AT^n$. (b) Enlarged plots for temperature dependence of resistance in HfGeTe$_4$ under pressure below 8 K. (c) Temperature dependence of resistance in HfGeTe$_4$ at 60.2 GPa under various magnetic fields up to 6 T. The inset shows the enlargement around the superconducting transition.

Figure 9 shows the pressure-phase diagram for (a) ZrGeTe$_4$ and (b) HfGeTe$_4$. Both samples exhibited the pressure-induced insulator to metal transition and superconductivity. Although the observed multi-step superconducting transitions in both isostructural materials ZrGeTe$_4$ and HfGeTe$_4$ are expected to be originated from a pressure-driven structural phase transitions, further investigations are required, for example, in-situ XRD analysis. The critical pressure for inducing superconductivity $P_c$ was lower and the $T_c$ was higher in HfGeTe$_4$ than those in ZrGeTe$_4$, possibly reflecting carrier doping from the Hf deficiency.

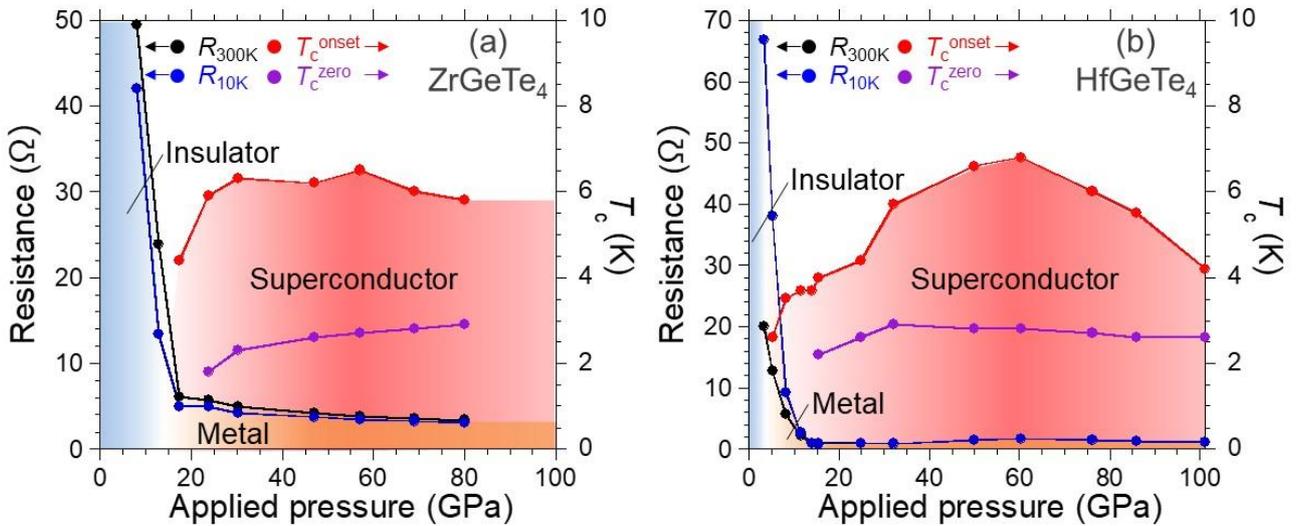

**Figure 9.** Pressure-phase diagram for (a) ZrGeTe$_4$ and (b) HfGeTe$_4$.

## 5. Conclusion

The Layered transition-metal chalcogenides (Zr,Hf)GeTe$_4$ were focused as candidate for pressure-induced superconductors via the data-driven approach using the database. The XRD and EDX analysis for the synthesized whisker-like single crystals revealed that only HfGeTe$_4$ shows Hf deficiency. The valence fluctuation in Zr and Hf in the crystals were suggested by the XPS analysis, which feature is consistent with the first report of the synthesis. The pressure-induced superconductivity in both materials were first time observed experimentally by using our originally designed DAC with the BDD electrodes and the UDD insulating layer. Since the investigated materials (Zr,Hf)GeTe$_4$ were recently focused as a superior layered compounds, it can be interested in an further investigation for a mechanism of the superconductivity as other TMDs.

**Acknowledgment**

The authors thank Dr. Hara for a discussion of the structural analysis. This work was partly supported by JST CREST Grant No. JPMJCR16Q6, JST-Mirai Program Grant Number JPMJMI17A2, JSPS KAKENHI Grant Number JP17J05926, 19H02177, and the "Materials research by Information Integration" Initiative (MI$^2$I) project of the Support Program for Starting Up Innovation Hub from JST. A part of the fabrication process of diamond electrodes was supported by NIMS Nanofabrication Platform in Nanotechnology Platform Project sponsored by the Ministry of Education, Culture, Sports, Science and Technology (MEXT), Japan. The part of the high-pressure experiments was supported by the Visiting Researcher's Program of Geodynamics Research Center, Ehime University. The computation in this study was performed on Numerical Materials Simulator at NIMS. The authors would like to acknowledge the ICYS Research Fellowship, NIMS, Japan.

**References**

[1] J. G. Bednorz and K. A. M¨uller, Z. Phys. B **64**, 189 (1986).

[2] Y. Kamihara, T. Watanabe, M. Hirano, and H. Hosono, J. Am. Chem. Soc. **130**, 3296 (2008).

[3] A.P. Drozdov, M.I. Eremets, I.A. Troyan, V. Ksenofontov, and S.I. Shylin, Nature **525**, 73 (2015).

[4] D. Duan, Y. Liu, F. Tian, D. Li, X. Huang, Z. Zhao, H. Yu, B. Liu, W. Tian, and T. Cui, Sci. Rep. **4**, 6968 (2014).

[5] M. Einaga, M. Sakata, T. Ishikawa, K. Shimizu, M. I. Eremets, A. P. Drozdov, I. A. Troyan, N. Hirao, and Y. Ohishi, Nature Phys. **12**, 835 (2016).

[6] S. Mozaffari, D. Sun, V. S. Minkov, A. P. Drozdov, D. Knyazev, J. B. Betts, M. Einaga, K. Shimizu, M. I. Eremets, L. Balicas, and F. F. Balakirev, Nat. Commun. **10**, 2522 (2019).

[7] W. Sano, T. Koretsune, T. Tadano, R. Akashi, and R. Arita, Phys. Rev. B **93**, 094525 (2016).

[8] A. P. Drozdov, P. P. Kong, V. S. Minkov, S. P. Besedin, M. A. Kuzovnikov, S. Mozaffari, L. Balicas, F. Balakirev, D. Graf, V. B. Prakapenka, E. Greenberg, D. A. Knyazev, M. Tkacz, M. I. Eremets, Nature **569**, (2019) 528–531.

[9] M. Somayazulu, M. Ahart, A. K. Mishra, Z. M. Geballe, M. Baldini, Y. Meng, V. V. Struzhkin, R. J. Hemley, Phys. Rev. Lett. **122**, 027001 (2019).

[10] H. Liu, I. I. Naumov, R. Hoffmann, N. W. Ashcroft, and R. J. Hemley, PNAS **114**,

6990–6995 (2017).

[11]  T. Ishikawa, T. Oda, N. Suzuki, and K. Shimizu, High Pressure Res. **35**, 37 (2015).

[12]  T. Ishikawa, A. Nakanishi, K. Shimizu, H. Katayama-Yoshida, T. Oda, and N. Suzuki, Sci. Rep. **6**, 23160 (2016).

[13]  S. Kiyohara, H. Oda, T. Miyata, and T. Mizoguchi, Sci. Adv. **2**, e1600746 (2016).

[14]  A. Seko, A. Togo, H. Hayashi, K. Tsuda, L. Chaput, and I. Tanaka, Phys. Rev. Lett. **115**, 205901 (2015).

[15]  T. Inoshita, S. Jeong, N. Hamada, and H. Hosono, Phys. Rev. X **4**, 031023 (2014).

[16]  Y. Hinuma, T. Hatakeyama, Y. Kumagai, L. A. Burton, H. Sato, Y. Muraba, S. Iimura, H. Hiramatsu, I. Tanaka, H. Hosono, and F. Oba, Nat. Commun. **7**, 11962 (2016).

[17]  R. Matsumoto, Z. Hou, H. Hara, S. Adachi, H. Takeya, T. Irifune, K. Terakura, and Y. Takano, Appl. Phys. Express **11**, 093101 (2018).

[18]  R. Matsumoto, Z. Hou, M. Nagao, S. Adachi, H. Hara, H. Tanaka, K. Nakamura, R. Murakami, S. Yamamoto, H. Takeya, T. Irifune, K. Terakura, and Y. Takano, Sci. Technol. Adv. Mater. **19**, 909-916 (2018).

[19]  R. Matsumoto, H. Hara, Z. Hou, S. Adachi, H. Tanaka, S. Yamamoto, Y. Saito, H. Takeya, T. Irifune, K. Terakura, and Y. Takano, Inorg. Chem. **59**, 325-331 (2020).

[20]  D. M. Rowe, CRC handbook of thermoelectrics, (CRC Press, 2010).

[21]  K. Kuroki and R. Arita, J. Phys. Soc. Jpn. **76**, 083707 (2007).

[22]  N. B. Kopnin, T. T. Heikkilä, and G. E. Volovik, Phys. Rev. B **83**, 220503 (2011).

[23]  S. G. Lie and J. P. Garbotte, Solid State Commun. **26**, 511-514 (1978).

[24]  A. A. Abrikosov, Physica C **341-348**, 97-102 (2000).

[25]  A. Mar and J. A. Ibers, J. Am. Chem. Soc. **114**, 3227-3238 (1993).

[26]  Y. Saito and J. Robertson, APL Materials **6**, 046104 (2018).

[27]  R. Matsumoto, Y. Sasama, M. Fujioka, T. Irifune, M. Tanaka, T. Yamaguchi, H. Takeya, and Y. Takano, Rev. Sci. Instrum. **87**, 076103 (2016).

[28]  R. Matsumoto, T. Irifune, M. Tanaka, H. Takeya, and Y. Takano, Jpn. J. Appl. Phys. **56**, 05FC01 (2017).

[29]  R. Matsumoto, A. Yamashita, H. Hara, T. Irifune, S. Adachi, H. Takeya, and Y. Takano, Appl. Phys. Express **11**, 053101 (2018).

[30]  R. Matsumoto, H. Hara, H. Tanaka, K. Nakamura, N. Kataoka, S. Yamamoto, T. Irifune, A. Yamashita, S. Adachi, H. Takeya, and Y. Takano, J Phys. Soc. Jpn. **87**, 124706 (2018).

[31]  G. M. Sheldrick, Acta Crystallogr., Sect. A **71**, 3 (2015).

[32]  L. J. Farrugia, J. Appl. Cryst. **45**, 849 (2012).

[33]  C. B. Hübschle, G. M. Sheldrick and B. Dittrich, J. Appl. Cryst. **44**, 1281 (2011).

[34]  K. Momma and F. Izumi, J. Appl. Cryst. **44**, 1272 (2011).

[35]  T. Miyayama, N. Sanada, M. Suzuki, J. S. Hammond, S.-Q. D. Si, and A. Takahara, J. Vac. Sci. Technol. A **28**, L1 (2010).

[36]  R. Matsumoto, Y. Nishizawa, N. Kataoka, H. Tanaka, H. Yoshikawa, S. Tanuma, and K. Yoshihara, J. Electron Spectrosc. Relat. Phenom. **207**, 55 (2016).

[37]  R. Matsumoto, Z. Hou, S. Adachi, M. Nagao, S. Yamamoto, P. Song, N. Kataoka, P. B.

Castro, K. Terashima, H. Takeya, H. Tanaka, T. Yokoya, T. Irifune, K. Terakura, and Y. Takano, High Press. Res. **40**, 22-34 (2020).

[38]　T. Irifune, A. Kurio, S. Sakamoto, T. Inoue, and H. Sumiya, Nature **421**, 599-600 (2003).

[39]　G. J. Piermarini, S. Block, J. D. Barnett, and R. A. Forman, J. Appl. Phys. **46**, 2774 (1975).

[40]　Y. Akahama and H. Kawamura, J. Appl. Phys. **96**, 3748 (2004).

[41]　G.J. Jang and H. Yun, Acta Cryst. **E64**, i27 (2008).

[42]　W. Gong, L. Li, P. Gong, Y. Zhou, Z. Zhang, W. Zhou, W. Wang, Z. Liu, and D. Tang, Appl. Phys. Lett. **114**, 172104 (2019).

[43]　Y. Nishino, A.R. Krauss, Y. Lin, and D.M. Gruen, J. Nucl. Mater. **228**, 346–353 (1996).

[44]　S. Lin and H. Li, Ceramics International **39**, 7677-7683 (2013).

[45]　X. Zhu, W. Ning, L. Li, L. Ling, R. Zhang, J. Zhang, K. Wang, Y. Liu, L. Pi, Y. Ma, H. Du, M. Tian, Y. Sun, C. Petrovic, and Y. Zhang, Sci. Rep. **6**, 26974 (2016).

[46]　S. J. Denholme, A. Yukawa, K. Tsumura, M. Nagao, R. Tamura, S. Watauchi, I. Tanaka, H. Takayanagi, and N. Miyakawa, Sci. Rep. **7**, 45217 (2017).

[47]　A. F. Kusmartseva, B. Sipos, H. Berger, L. Forró, and E. Tutiš, Phys. Rev. Lett. **103**, 236401 (2009).

[48]　H. F. Zhai, Z. T. Tang, H. Jiang, K. Xu, K. Zhang, P. Zhang, J. K. Bao, Y. L. Sun, W. H. Jiao, I. Nowik, I. Felner, Y. K. Li, X. F. Xu, Q. Tao, C. M. Feng, Z. A. Xu, and G. H. Cao, Phys. Rev. B **90**, 064518 (2014).

[49]　R. Singha, A. K. Pariari, B. Satpati, and P. Mandal, PNAS **114**, 2468–2473 (2017).

[50]　M. Naito and S. Tanaka, J. Phys. Soc. Jpn. **51**, 219-227 (1982).

[51]　F. F. Tafti, Q. D. Gibson, S. K. Kushwaha, N. Haldolaarachchige, and R. J. Cava, Nat. Phys. **12**, 272–277 (2016).

[52]　S. Sun, Q. Wang, P. Guo, K. Liu, and H. Lei, New J. Phys. **18**, 082002 (2016).